%% file: triso.tex
\documentclass[12pt,onecolumn]{IEEEtran}
\include{package}

\include{defs}

\pdfoutput=1

\title{Model-based Reconstruction for Enhanced X-ray CT of Tri-structural Isotropic (TRISO) Particles}

\author{Singanallur~V.~Venkatakrishnan$^{\star}$,
  Amirkoushyar Ziabari$^{\star}$,
  Philip Bingham$^{\star}$,
  Grant Helmreich $^{o}$
  \thanks {$^{\star}$ Multimodal Sensor Analytics Group, Oak Ridge National Lab, Oak Ridge, TN 37831, USA}
  \thanks{$^{o}$ Nuclear Fuels Group, Oak Ridge National Lab, Oak Ridge, TN 37831, USA}
 \thanks{\scriptsize{This manuscript has been authored by UT-Battelle, LLC, under Contract No. DE-AC05-00OR22725 with the U.S. Department of Energy.
The United States Government and the publisher, by accepting the article for publication, acknowledges that the United States Government retains a non-exclusive, paid-up, irrevocable, world-wide license to publish or reproduce the published form of this manuscript, or allow others to do so, for United States Government purposes.
DOE will provide public access to these results of federally sponsored research in accordance with the DOE Public Access Plan (http://energy.gov/downloads/doe-public-access-plan).
 }}
}

\begin{document}

\maketitle
\thispagestyle{empty}
\pagestyle{empty}

\input{front}

\input{trisoBody}

\bibliographystyle{IEEEbib}
\footnotesize
\bibliography{triso}


\end{document}

%% file: package.tex
\usepackage{array}
\usepackage{mdwmath}
\usepackage{mdwtab}
\usepackage{fixltx2e}
\usepackage{url}
\usepackage{color}
\usepackage[]{graphicx}
\usepackage{epsfig}
\usepackage{bbm}
\usepackage{enumerate}
\usepackage{amsfonts}
\usepackage{amsmath}
\usepackage{amssymb}
\usepackage{cite}
\usepackage{algorithm}
\usepackage{algpseudocode}
\usepackage{eqparbox}
\usepackage{comment}
\usepackage{breqn}
\usepackage{dblfloatfix}
\usepackage{todonotes}

%% file: defs.tex
\providecommand{\norm}[1]{\lVert#1\rVert}

\DeclareMathOperator*{\argmin}{argmin}

%% file: front.tex
\begin{abstract}
  Tri-Structural Isotropic (TRISO) fuel particles are a key component of next generation nuclear fuels. 
  Using X-ray computed tomography (CT) to characterize TRISO particles is challenging because of the strong attenuation of the X-ray beam by the uranium core leading to severe photon starvation in a substantial fraction of the measurements.
  Furthermore, the overall acquisition time for a high-resolution CT scan can be very long when using conventional lab-based X-ray systems and reconstruction algorithms.   
  Specifically, when analytic methods like the Feldkamp-Davis-Kress (FDK) algorithm is used for reconstruction, it results in severe streaks artifacts and noise in the corresponding 3D volume which make subsequent analysis of the particles challenging. 
  In this article, we develop and apply model-based image reconstruction (MBIR) algorithms for improving the quality of CT reconstructions for TRISO particles in order to facilitate better characterization.
  We demonstrate that the proposed MBIR algorithms can significantly suppress artifacts with minimal pre-processing compared to the conventional approaches. 
  Furthermore, we demonstrate the proposed MBIR approach can obtain high-quality reconstruction compared to the FDK approach even when using a fraction of the typically acquired measurements, thereby enabling dramatically faster measurement times for TRISO particles. 
\end{abstract}

%% file: trisoBody.tex
\section{Introduction}

\begin{figure*}
\begin{center}
    \includegraphics[scale=0.45,trim=0cm 7.5cm 1cm 0cm,clip]{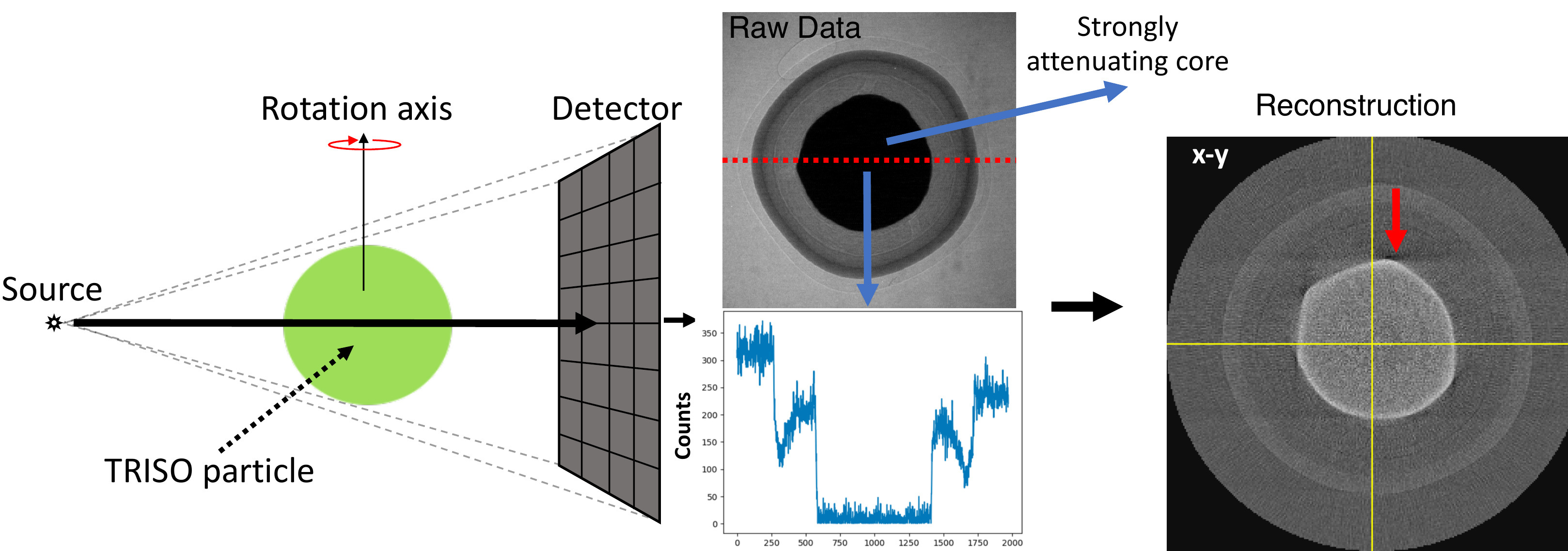}
\end{center}
\caption{\label{fig:cbct_setup} Illustration of the CT scan process for TRISO particles. 
The sample is rotated about a single axis in a cone-beam CT scanner and at each position a single projection image is acquired using a high-resolution detector. 
Due to the strongly attenuating uranium core, we get measurements with very few counts at the center of the detector as shown in the image above (displayed in the range of 0-350 counts), resulting in strong artifacts in reconstructions (streaks indicated using arrow, noise etc.) obtained using conventional algorithms such as the FDK \cite{feldkamp1984practical} method. 
}
\end{figure*}
Tri-structural isotropic (TRISO) particles are a promising new fuel technology for next-generation nuclear reactors.
These particles are approximately spherical, about 1 mm in diameter and are composed of a uranium based kernel surrounded by several layers of carbon and silicon carbide cladding designed to retain the uranium kernel and fission products generated in a reactor. 
Characterization of the cladding regions of TRISO particles is vital in order to better understand their structure and how they degrade in the course of operation of a nuclear reactor.
One standard approach for characterization is focused ion-beam scanning electron microscopy (FIB-SEM) \cite{jose_fibsem} imaging in which the sample is imaged one layer at a time by polishing off a layer of the material. 
In contrast to FIB-SEM, X-ray computed tomography (CT) enables \textit{non-destructive characterization (NDC)} of TRISO particles \cite{lowe_triso,helmreich2020method} and hence is gaining popularity as the method of choice for NDC. 
X-ray CT is typically carried out by illuminating a sample with a poly-chromatic X-ray source and measuring the attenuated beam using a high-resolution area detector (see Fig.~\ref{fig:cbct_setup}).
Several such projection measurements are obtained by rotating the sample about a single axis, after which a cone-beam CT (CBCT) reconstruction algorithm such as the Feldkamp-Davis-Kress (FDK) \cite{feldkamp1984practical} is used to obtain a 3D reconstruction corresponding to the sample.

While XCT is a well established characterization technique, the imaging of TRISO particles poses several challenges.
First, these particles strongly attenuate the X-ray beam due to the presence of a uranium core resulting in very few counts in a large region (see Fig.~\ref{fig:cbct_setup}) of the detector.
If such data is directly used for reconstructions, it can results in significant artifacts (see Fig.~\ref{fig:cbct_setup}) which hamper the use of image analysis algorithms to perform accurate characterization of the cladding regions of the particle. 
We note that while some of these artifacts are similar to the \textit{metal artifacts} observed in medical and baggage-scan X-ray CT, which are caused by beam-hardening\cite{katsura2018current}, the primary cause of artifacts in the imaging of TRISO particles are due to photon-starvation. 
Thus the conventional metal artifact reduction (MAR) algorithms \cite{gjesteby2016metal} are not directly applicable to this case.  
Furthermore, for typical exposure settings and X-ray cone-beam CT (CBCT) scanners used, it can take a very large number of projection measurements to get a reasonable reconstruction quality when the FDK algorithm is used. 
Because of the long duration required to measure a single projection image, this implies that the overall measurement time, consisting of thousands of such images, is of the order of one day to complete a typical scan. 
In summary, the existing measurement protocols and algorithms used for XCT of TRISO particles result in a characterization system with long measurement times and sub optimal reconstruction quality. 

In this paper, we present algorithms for improving the quality of CT reconstructions for TRISO particles.
We use model-based image reconstruction (MBIR) algorithms \cite{Bo641Text} as the basis for improving reconstruction quality while also enabling significant reduction in the measurement time for these particles. 
MBIR approaches have been widely developed for various CT applications demonstrating that it is possible to reduce dose \cite{ZhouNHICD}, shorten the measurement time \cite{AdityaICASSP14} and get high-quality reconstructions from limited view data \cite{VenkatHAADF13}. 
MBIR approaches involve formulating the reconstruction task as minimizing a high dimensional cost function that balances two sets of terms - one that incorporates the physics of the image formation and noise statistics of the detector; and a second term that accounts for a model for the sample to be imaged. 
Our core contribution is the modification of the noise modeling term in the MBIR approach to account for the highly-attenuated regions of the measurements and effectively reconstruct the sample using only measurements which are reliable.  
This can be easily done in the MBIR framework by setting a weight term based on the actual measured signal and performing the optimization on a sub-set of the measured pixels. 
 Using experimental data,
  we demonstrate that the MBIR method can dramatically improve upon the FDK algorithm by suppressing artifacts and noise in the reconstruction.
Furthermore, we also highlight the ability of the MBIR method to obtain high-quality reconstructions using a fraction of the typically measured data, illustrating that it is possible to dramatically accelerate the measurement process while preserving the quality of the reconstructed images. 
 
\section{Model-based Image Reconstruction}
\label{sec:mbir}

In order to reconstruct the samples in 3D from the raw measurements, 
we use the MBIR \cite{ZhouNHICD} framework.
The reconstruction is formulated as a minimization problem, 
\begin{eqnarray}
\label{eq:MBIRCost}
\hat{f} \leftarrow \argmin_{f} \left\{ l(g;f) + s(f)\right\}
\end{eqnarray}
where $g$ is the vector of projection measurements, $f$ is the 
vector containing all the voxels,
$l(;)$ is a data fidelity enforcing function 
and $s(.)$ is a function that enforces regularity in $f$. 
We propose to use the well-established quadratic data-fidelity term \cite{SaBo92,SaBo93} of the form 
\begin{eqnarray}
  l(g;f) = \frac{1}{2}\norm{g-Af}_{W}^{2}
  \label{eq:LLD}
\end{eqnarray}
where $A$ is a forward projection matrix that accounts for the the cone-beam geometry, and $W$ is a diagonal matrix with entries set to be the inverse variance of the noise in $g$, and $g$ is a vector containing the log-normalized projection measurements.
We design $A$ to model the cone-beam geometry 
by using the ASTRA tool-box \cite{AstraGPU11,AstraUltramic15}
that can utilize multiple GPUs \cite{vanAarleASTRA16,BleichrodtAstra16}
to accelerate the application of this matrix.
However, the projection ($A$) and back-projection ($A^{T}$) matrices
are not perfectly matched.

For $s(f)$, we choose the negative log of
q-generalized Markov-random field (qGGMRF)
probability density function \cite{JBSaBoHsMultiSlice}. 
It is given by
\begin{eqnarray}
\label{eq:Prior}
s(f)&=& \sum\limits _{\{j,k\}\in \mathcal{N}}w_{jk}\rho(f_j-f_k) \nonumber \\
\rho(f_j-f_k )&=&\frac{\left|\frac{f_{j}-f_{k}}{\sigma_f}\right|^{2}}{c + 
\left|\frac{f_{j}-f_{k}}{\sigma_f}\right|^{2-p}} \nonumber
\end{eqnarray}
$\mathcal{N}$ is the set 
of pairs of neighboring voxels (e.g. a 26 point neighborhood),
$1\leq p \leq 2$, $c$ and $\sigma_f$ are
qGGMRF parameters. 
The weights $w_{jk}$ are 
inversely proportional to the distance 
between voxels $j$ and $k$, normalized to $1$. 
This model provides a greater degree of flexibility in the quality of reconstructions compared to an algorithm specifically designed for a total-variation regularizer that may force the reconstructions to appear ``waxy'' \cite{Bo641Text}.
Combining the data fidelity model \eqref{eq:LLD} 
with the image model \eqref{eq:Prior} the MBIR cost function is 
\begin{dmath}
\label{eq:OriginalCost}
c(f)=\frac{1}{2}  \norm{g-Af}_{W}^{2} + s(f) 
\end{dmath}
Thus, the reconstruction is obtained by
\begin{eqnarray*}
\hat{f} \leftarrow \argmin_{f} c(f)
\end{eqnarray*}

\section{Modified MBIR for TRISO Measurements}
\label{sec:proposed_algo}

\begin{figure}[!h]
\begin{center}
    FDK  \hspace{2.5in}MBIR \\
\begin{tabular}{l}
    \includegraphics[scale=0.5,trim=0cm 1cm 0cm 1cm,clip]{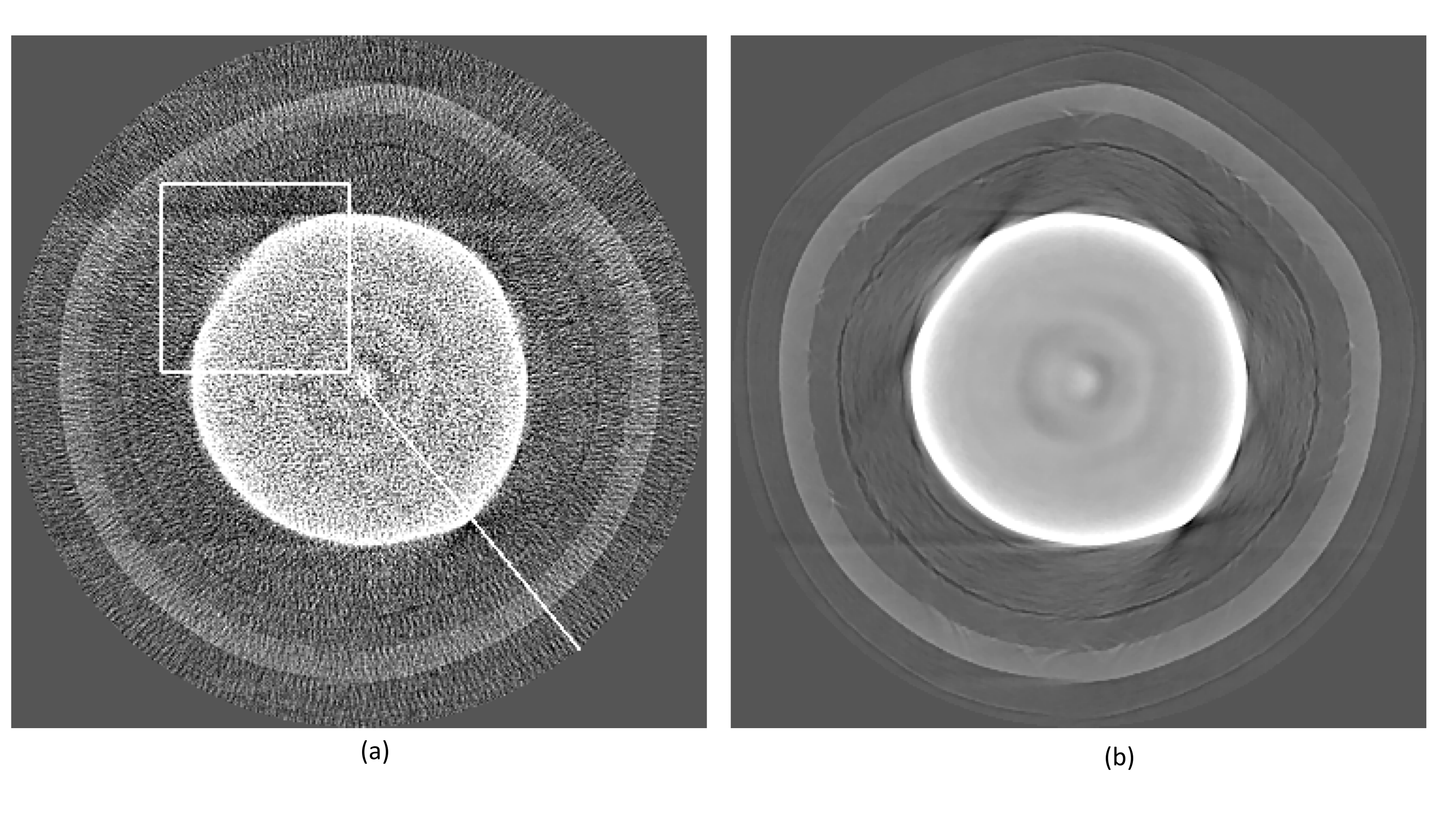} 
    \tabularnewline
        \includegraphics[scale=0.5,trim=0cm 8.25cm 0cm 0cm,clip]{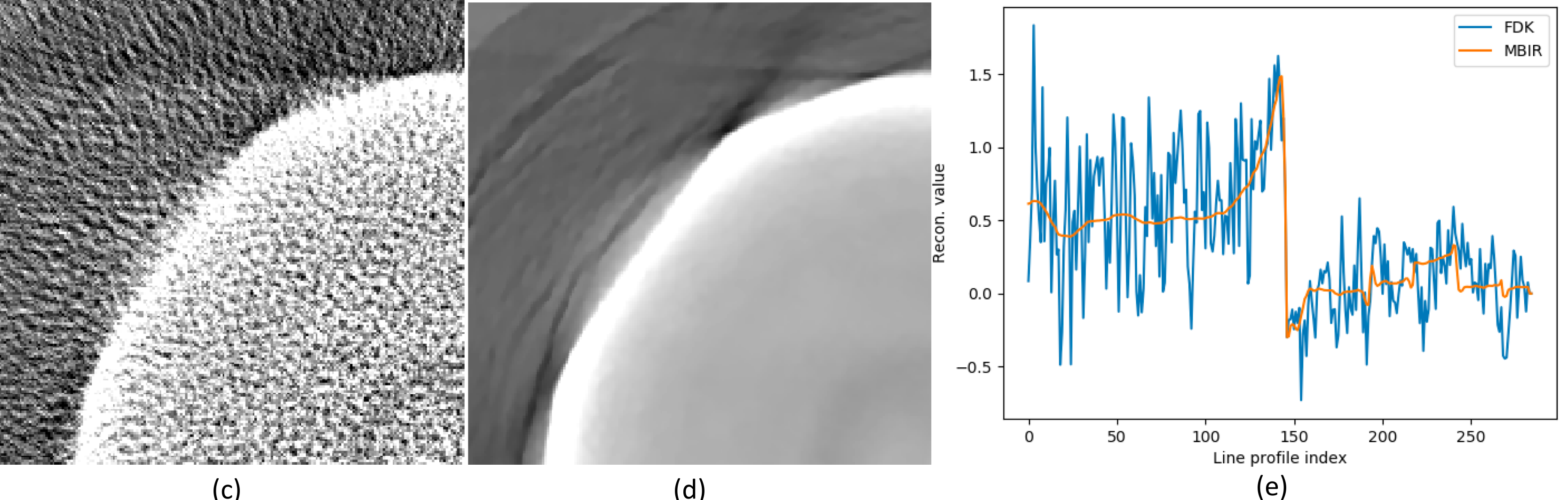}
\end{tabular}
\end{center}
\caption{\label{fig:fdkvsmbir} Cross-section from the (a) FDK  and (b) conventional MBIR reconstruction without any special pre-processing of the data. 
(c) and (d) are patches from (a) and (b) for a more detailed view. 
Notice that the MBIR reconstructions have much lower noise and artifacts compared to the FDK reconstruction due to the photon starvation in large regions of the original measurements. 
(e) Shows a line profile indicated in (a) from the FDK and MBIR reconstruction highlighting the various artifacts in the reconstruction. 
}
\end{figure}
\begin{figure}[!h]
\begin{center}
     FDK  \hspace{2.5in} MBIR \\
\begin{tabular}{l}
    \includegraphics[scale=0.5,trim=0cm 1cm 0cm 1cm,clip]{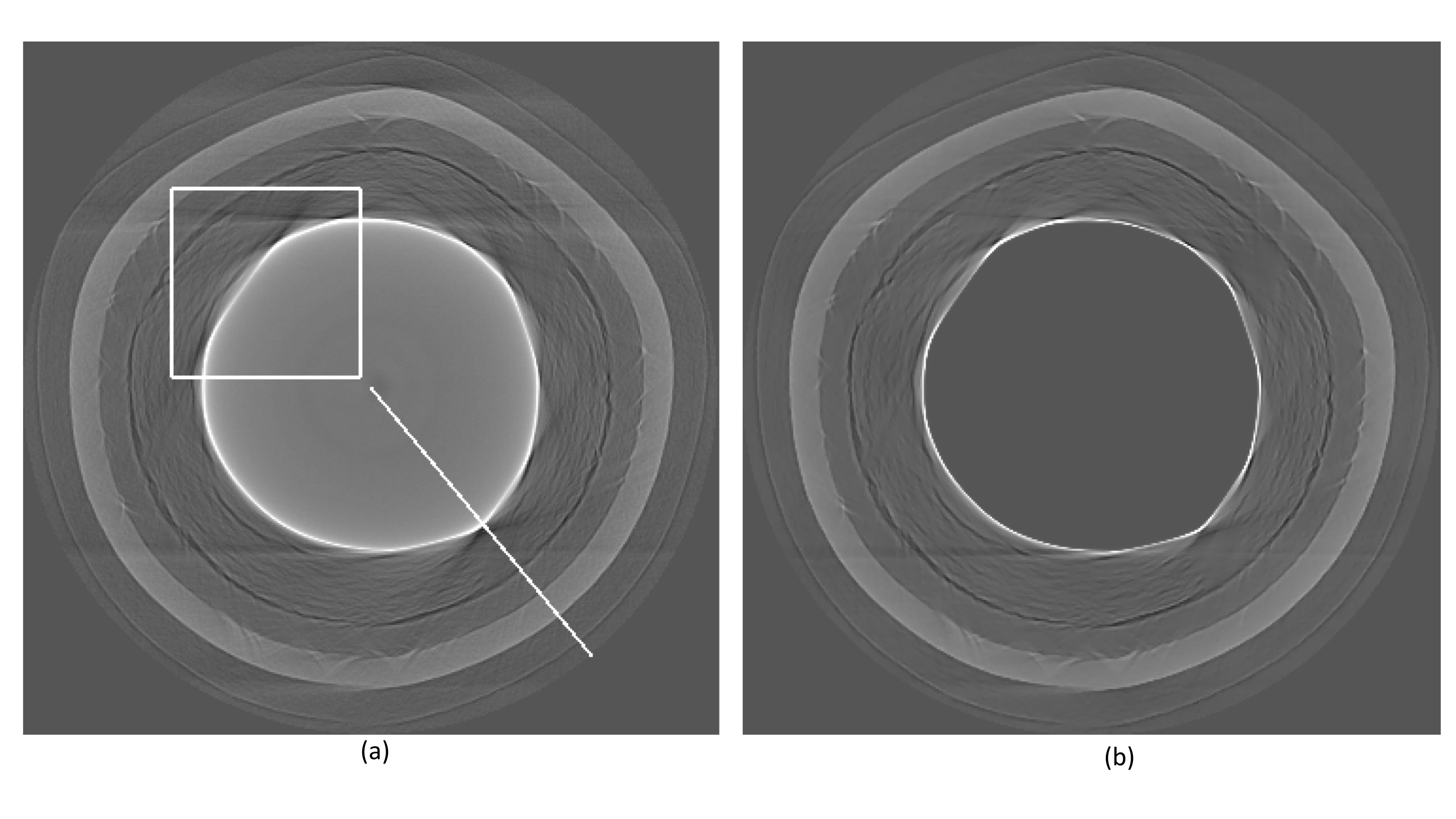} \\
        \includegraphics[scale=0.5,trim=0cm 8.5cm 0cm 0cm,clip]{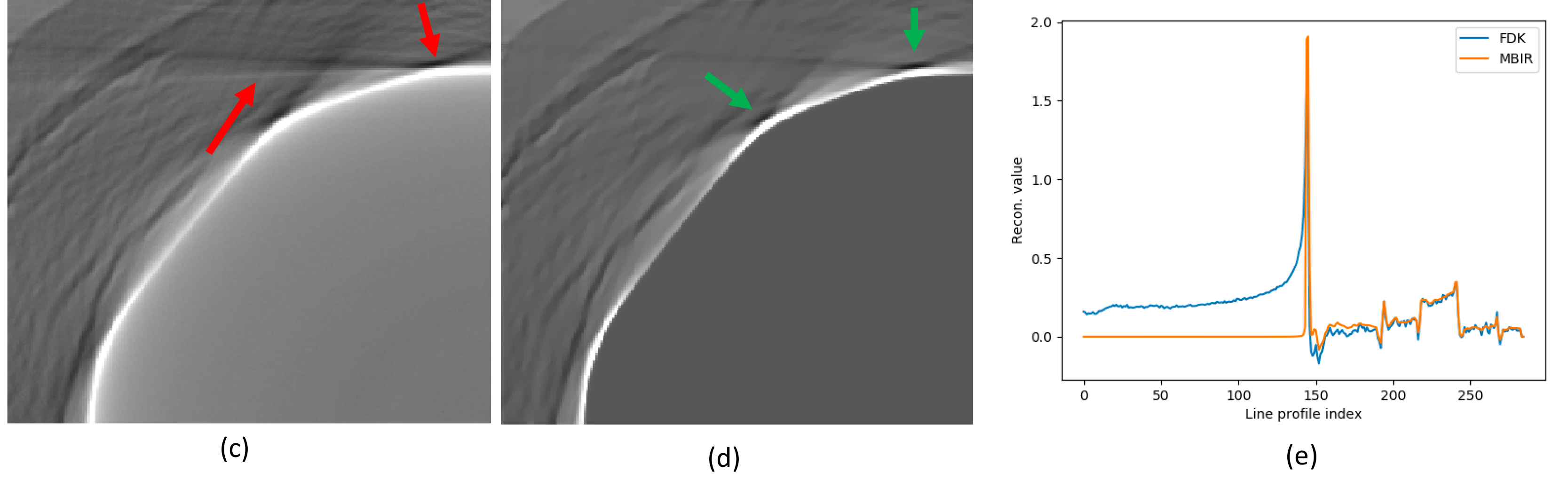} \\
  \end{tabular}
\end{center}
\caption{\label{fig:fdkvsmbir2} Cross-section from the (a) FDK  and (b) the proposed MBIR reconstruction. 
The FDK reconstruction is obtained by pre-processing the original count data by clipping the raw count data to a minimum value to reduce the influence of the noisy low-count measurements.   
(c) and (d) are patches from (a) and (b) for a more detailed view. 
Notice that the proposed MBIR reconstruction dramatically suppresses noise and artifacts compared to the FDK reconstruction as well as the MBIR reconstruction (marked in green arrows) in Fig.~\ref{fig:fdkvsmbir}. 
(e) Shows a line profile indicated in (a) from the FDK and MBIR reconstruction highlighting the various artifacts in the reconstruction.}
\vspace{-0.1in}
\end{figure}

One of the main challenges in obtaining high quality reconstructions for X-ray CT of TRISO particles is the strong attenuation of X-rays by the core.
For example, Fig.~\ref{fig:cbct_setup} shows a line profile from a single projection image acquired from the X-ray CT scan of a TRISO particle.
Notice that the counts in a central region rapidly drops to almost zero due to the beam being blocked by the uranium core, making these measurements extremely unreliable to use in order to do reconstructions. 
If the conventional FDK algorithm is directly used to process such data, it can result in significant artifacts (see Fig.~\ref{fig:cbct_setup}). 

One way to address this challenge is to directly use the MBIR algorithm \eqref{eq:MBIRCost}. 
Since the weight term in the MBIR framework is typically set such that $W_{ii}=\lambda_i$, where $\lambda_i$ is the raw measurement count \cite{SaBo92}, we could expect this to help in reducing artifacts because the cost-function terms corresponding to very low counts are naturally weighted less compared to the other terms. 
Empirically, we observe (see section \ref{sec:results}) that this results in a reduction of artifacts but we still continue to see some ``blooming'' and streak artifacts around the corners of the core because of the inherent noise in the low-count data. 
Instead, we propose to simply set this weight to zero if the detected counts are below a certain threshold i.e. 
\begin{eqnarray}
\tilde{W}_{ii}= \left\{
\begin{array}{ll}
      \lambda_{i} & \lambda_{i} \geq T \\
      0 & \lambda_{i} < T \\
\end{array} 
\right.
\label{eq:new_weight}
\end{eqnarray}
where $T$ is a pre-determined threshold. 
This step effectively has the impact of completely leaving out those measurements which are deemed unreliable.
With this simple modification, the resulting cost function to optimize using the MBIR framework is
\begin{dmath}
\label{eq:NewCost}
\tilde{c}(f)=\frac{1}{2}  \norm{g-Af}_{\tilde{W}}^{2} + s(f). 
\end{dmath}
Thus, the reconstruction is obtained by
\begin{eqnarray*}
\hat{f} \leftarrow \argmin_{f} \tilde{c}(f)
\end{eqnarray*}
We emphasize that this approach cannot be directly extended to the FDK algorithm because it only operates on the normalized projection data, and hence there is no straight-forward means to leave out the low-count projection data as a part of the reconstruction. 

We use the optimized gradient method (OGM) \cite{KimOGM15}
to find a minimum of the cost function in \eqref{eq:NewCost}.
The algorithm involves a standard gradient computation
combined with a step-size determined using Nesterov's method.
Specifically, for each iteration $k$, 
\begin{eqnarray}
  h^{(k+1)}  \leftarrow  f^{(k)} - \frac{1}{L} \nabla \tilde{c}(f^{(k)}) \\
  t^{(k+1)}  \leftarrow  \frac{1+\sqrt{1+4(t^{(k)})^2}}{2} \\
  f^{(k+1)}  \leftarrow  h^{(k+1)} + \frac{t^{(k)}-1}{t^{(k+1)}}(h^{(k+1)}-h^{(k)})  \nonumber \\   + \frac{t^{(k)}}{t^{(k+1)}}(h^{(k+1)}-f^{(k)})
  \label{eq:ogm_update}
\end{eqnarray}
where $t^{(0)}=1$, $L$ is the Lipschitz constant of the gradient of $\tilde{c}(.)$, $h^{(0)}=f^{(0)}$ is an initial estimate for the reconstruction.

The gradient of the cost-function $c(.)$ is given by
\begin{eqnarray}
  \nabla \tilde{c}(f) = -A^{T}\tilde{W}(g-Af) + \nabla s(f).
\end{eqnarray}

\begin{figure*}[!htbp]
\includegraphics[scale=0.515,trim=0cm 1.32cm 0cm 0cm,clip]{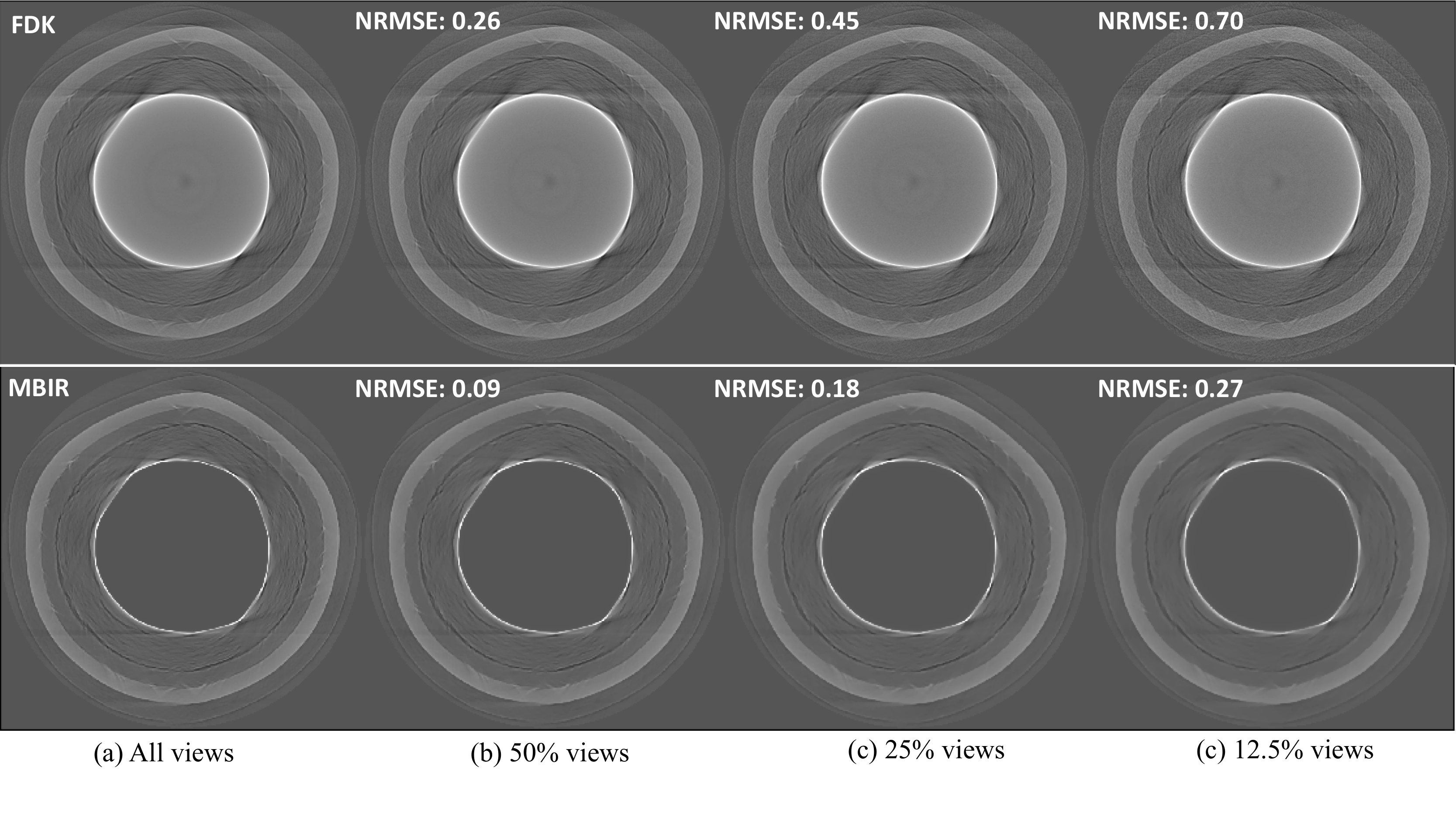}
\caption{\label{fig:fdkvsmbir_vieweffect} Cross-section from the 3D volume reconstructed using the modified FDK and the proposed MBIR approach as a function of the number of projection images.
Notice that the noise in the FDK reconstruction increases significantly as the number of projection data used is decreased.
In contrast the MBIR reconstruction are of higher quality even when only a fraction of the original measurements are used, indicating that it is possible to reduce the measurement time significantly while preserving quality. 
The NRMSE between the sparse view and full-view reconstruction (in percentage) is also shown highlighting the slower degradation in performance of the MBIR technique compared to the FDK method.
\vspace{-0.1in}
}
\end{figure*}
\section{Results}
\label{sec:results}
We measure a single TRISO particle using an Xradia MicroXCT-400 X-ray CT system with a Hamamatsu source operated at 40kV and 8W along with a detector of size $1972 \times 1972$ pixels. 
3200 projection images were acquired with an exposure time of 15 seconds plus some time for processing and repositioning, resulting in a total acquisition time of approximately 20 hours.
The system automatically shifts the detector for each acquisition image in order to be able to deal with ring-artifacts \cite{pelt2018ring}. 
We begin by pre-processing the data-set by applying a median filter with window size set to $7$ in order to suppress the effect of impulse noise due to gamma/X-ray strikes on the detector which can cause streaks in the reconstruction.
We then normalize the data using the open-beam measurements followed by applying a shift to each (normalized) projection image to account for the detector shifts during acquisition. 
We then apply different reconstruction algorithms to this data sets in order to compare their performance. 
Since we do not have access to the ground-truth we rely on the visual quality of the reconstructions for the comparisons. 

First, we compare the performance of the FDK and MBIR methods when no additional processing is done to account for the severe photon starvation in the data.  
Fig.~\ref{fig:fdkvsmbir} shows a single cross-section from the reconstructed volume using the FDK and MBIR algorithms. 
Notice that the FDK reconstruction has severe noise in the cross-section because of the noisy and photon-starved measurements. 
The MBIR algorithm significantly improves the reconstruction compared to the FDK method because of the natural weighting in the cost function because of an accurate noise model. 
Next, we compare the proposed MBIR algorithm to an algorithm that involves applying pre-processing to the data followed by the FDK algorithm. 
For the FDK method, in regions where the weights of the original data are less than $50$ counts, we clip the weights to be exactly $50$. 
In the case of the MBIR method, the threshold $T$ in equation \eqref{eq:new_weight} is set to $50$ so that measurements corresponding to this value are ``rejected'' in the reconstruction.
Fig.~\ref{fig:fdkvsmbir2} shows a single cross-section and a line profile through the sample for each of these methods.
Notice that the FDK method results in significant improvements compared to the results with no pre-processing (Fig.~\ref{fig:fdkvsmbir}), but still has strong streaks and noise. 
The MBIR method suppresses these artifacts compared the FDK. 
Furthermore compared to the baseline MBIR method, we observe that the dark spots around the corners are less smeared out, enabling better characterization of the cladding regions around the core. 
In summary, the proposed MBIR method leads to significant improvements in performance with a small modification to the original MBIR algorithm.

Finally, we compare the performance of the improved FDK and proposed MBIR method as a function of number of projection images used for reconstruction; using 3200 (original), 2400, 1600, 800 images (corresponding to a sub-sampling rate of 50\%, 25\%, and 12.5\% respectively) and comparing the reconstruction performance. 
Fig.~\ref{fig:fdkvsmbir_vieweffect} shows a single cross-sections from the reconstruction for the different sub-sampling rate along with the normalized root mean squared error (NRMSE) between the full-view and sparse-view reconstruction.  
Notice that the FDK method starts to significantly degrade in performance, while the MBIR preserves the performance as the sub-sampling rate is increased, highlighting how we can perform the measurements faster while preserving the structural details. 
Conservatively, even 1600 measurements result in sufficiently high quality reconstructions, suggesting that it may be possible to reduce the acquisition time by at least $50\%$ while preserving the image quality in the reconstruction when the proposed MBIR algorithm is used. 

\section{Conclusion}
\label{sec:concl}
In this paper, we presented a model-based reconstruction algorithm for improving CT reconstruction quality of core-shell TRISO particles which are a promising technology for nuclear fuels. 
We demonstrated that despite of the severe photon starvation in the measurements due to the dense core, the proposed MBIR algorithm can enable higher quality reconstruction of the buffer regions compared to the FDK method that is typically used. 
We also demonstrate that it is possible to dramatically accelerate the scan time for these particles while preserving the details in the reconstruction.